\documentclass[sn-mathphys-num]{sn-jnl}

\usepackage{graphicx}%
\usepackage{multirow}%
\usepackage{amsmath,amssymb,amsfonts}%
\usepackage{amsthm}%
\usepackage{mathrsfs}%
\usepackage[title]{appendix}%
\usepackage{subcaption}
\usepackage{siunitx,booktabs}

\usepackage{xcolor}%
\usepackage{textcomp}%
\usepackage{manyfoot}%
\usepackage{booktabs}%
\usepackage{algorithm}%
\usepackage{algorithmicx}%
\usepackage{algpseudocode}%
\usepackage{listings}%
\theoremstyle{thmstyleone}%
\theoremstyle{thmstyletwo}%

\theoremstyle{thmstylethree}%

\begin{document}

\title[Article Title]{Impact of HRTF individualisation and head movements in a real/virtual localisation task}


\author*[1,2]{\fnm{Vincent} \sur{Martin}}\email{vincentyong.martin@entpe.fr}

\author[1]{\fnm{Lorenzo} \sur{Picinali}}\email{l.picinali@imperial.ac.uk}

\affil[1]{\orgname{Imperial College London}, \orgdiv{Dyson School of Design Engineering}, \orgaddress{\city{London}, \postcode{SW7 2BD}, \country{United Kingdom}}}
\affil[2]{\orgname{ENTPE}, \orgdiv{CNRS, LTDS, UMR5513}, \orgaddress{\city{Vaulx-en-Velin}, \postcode{69518}, \country{France}}}


\abstract{The objective of Audio Augmented Reality (AAR) applications are to seamlessly integrate virtual sound sources within a real environment. It is critical for these applications that virtual sources are localised precisely at the intended position, and that the acoustic environments are accurately matched.

One effective method for spatialising sound on headphones is through Head-Related Transfer Functions (HRTFs). These characterise how the physical features of a listener modify sound waves before they reach the eardrum. This study examines the influence of using individualised HRTFs on the localisation and the perceived realism of virtual sound sources associated with a real visual object. 

Participants were tasked with localising virtual and real speech sources presented via headphones and through a spherical loudspeaker array, respectively. The assessment focussed on perceived realism and sources location. All sources were associated with one of thirty real visual sources (loudspeakers) arranged in a semi-anechoic room.

Various sound source renderings were compared, including single loudspeaker rendering and binaural rendering with individualised or non-individualised HRTFs. Additionally, the impact of head movements was explored: ten participants completed the same task with and without the possibility to move their head.

The results showed that using individual HRTFs improved perceived realism but not localisation performance in the static scenario. Surprisingly, the opposite was observed when head movements were possible and encouraged.}




\maketitle

\section{Introduction}\label{sec1}

Immersive audio aims to create a realistic and three-dimensional sound experience from the perspective of a listener. Rendering technologies simulate how sounds originate from various directions and distances, enhancing the feeling of presence and realism. With the emergence of affordable consumer technologies, immersive audio has become more important, driving a significant increase in the development and adoption of Audio Augmented Reality (AAR) applications. In AAR, two critical features must be reproduced during the sound rendering. First, sound sources must be perceived as located at the intended positions within the user's surrounding environment. Second, the sound rendering must achieve a level of realism that allows it to be seamlessly integrated within this environment. 

Most AR applications rely on headphones, enabled by the use of binaural rendering. This creates a three-dimensional auditory experience over headphones by simulating how sound waves interact with the human ears, head, shoulders and torso \cite{blauert1997spatial, majdak2014acoustic}. This interaction, described by the Head-Related Transfer Function (HRTF), varies among individuals due to their unique morphological characteristics. For a faithful virtual reproduction of spatialised sound sources, individual HRTFs should be used. However, acoustically measuring these HRTFs is time-consuming, technically challenging, and expensive to implement \cite{engel2023sonicom}. Consequently, binaural rendering often employs non-individualised (or ``generic'') HRTFs, or partially individualised ones \cite{picinali2023system}. It has been shown that using non-individual HRTFs fail at conveying accurate localisation along a particular cone of confusion \cite{wenzel1993localization}, as well as spatial release from masking abilities when trying to understand speech in noise \cite{cuevas2021impact, gonzalez2024spatial}. 

Additionally, individual HRTFs generally contribute to a more realistic perception. Realism is achieved when physical cues create a unified impression that aligns with a listener's experiences and beliefs about the world. This is heavily related to externalisation \cite{best2020sound}, which refers to the perception that a sound source is outside the head, and can be considered as a prerequisite for realistic binaural rendering.
Research demonstrates that the use of individual HRTFs enhances the externalisation of virtual sound sources \cite{werner2016summary}, but this effect is less consistent for speech stimuli \cite{moller1996binaural, begault2001direct} and their perceived realism \cite{usher2007perceived}. Several studies indicate that acoustically measured HRTFs contain perceptually irrelevant details for localisation performance \cite{kulkarni1998role,kulkarni1999sensitivity,breebaart2001perceptual}. The relevance of individualisation may vary depending on the context in which virtual sound sources are perceived. Two aspects are crucial for evaluating the perceptual characteristics of binaurally rendered stimuli: the presence of head movements and the influence of vision on auditory perception.

Regarding head movements, it has been shown that head-tracked binaural rendering improves localisation performance for real sound sources \cite{gaveau2022benefits}, and in resolving front-back confusions often caused by non-individual HRTFs \cite{wallach1940role, wightman1999resolution, brimijoin2012role}. Head movements also enhance the externalisation of stimuli if the movements are sufficiently large \cite{hendrickx2017influence}. Some studies suggest that the benefits of head movements can surpass those induced by HRTF individualisation. For example, Begault et al. \cite{begault2001direct} found that individual HRTFs did not lead to better localisation performance or externalisation if compared to non-individual HRTFs for speech stimuli presented in the horizontal plane when participants' head movements were tracked. Similar findings were reported by Oberem et al. \cite{oberem2020experiments}, who demonstrated that head movements improve localisation accuracy and externalisation, while reducing front–back reversals. They also noted that the pronounced benefits of head movements may mask the potential effects of HRTF individualisation. 

In AAR contexts, virtual sound sources can be linked to visual cues, creating an auditory-visual bond. Multisensory integration research shows how visual perception influences the auditory one \cite{howard1966human}. For example, it has been speculated that a contributing factor to externalisation and realism is the presence of congruent visual information supporting the existence of an externalised sound source \cite{werner2016summary}. Another example is the ventriloquist effect where the perceived location of a sound source shifts towards a visual object \cite{bertelson1998automatic}. This plasticity in multisensory spatial integration, combined with the higher spatial accuracy of vision over audition \cite{king2009visual}, is significant for AR applications associating virtual sound sources with real visual objects. Consequently, even if a virtual sound source's spatial rendering is inaccurate, it can be perceived at the intended position occupied by a real visual object \cite{heller2014simplifying, kyto2015ventriloquist}. 

The combined effects of HRTF individualisation, visual cues, and room reverberation reproduction in the presence of head movements were examined by Starz et al. \cite{starz2025comparison}, who found no significant contribution of HRTF individualisation compared to the other factors. Notably, participants in their study wore a head-mounted display that did not provide direct visual cues linked to the simulated sound source. The authenticity of virtual sound sources in an augmented acoustic reality (AAR) context has been investigated more specifically by Brinkmann et al. \cite{brinkmann2017authenticity}, using individual HRTFs in both reverberant and anechoic conditions. They reported that head movements reduced authenticity, leading to more frequent identification of virtual stimuli as virtual.

These different studies suggest that while individualised HRTFs improve localisation and realism, their benefits can be surpassed by the presence of head movements. Additionally, visual cues further enhance localisation and realism. Therefore, it can be hypothesised that in an AAR context, the presence of both head movements and visual cues might negate the need for individualised HRTFs from the listener's standpoint. An important distinction must be made when evaluating the realism of a virtual sound source: it can be assessed in terms of ``authenticity" or ``plausibility". Oberem et al. \cite{oberem2016experiments} showed that, in static and anechoic conditions with speech stimuli, both “authenticity” (i.e., indistinguishability of a binaural stimuli from a loudspeaker-produced stimulus) and “plausibility” (i.e., perceived as a real sound source) can be achieved with individual HRTFs. Lindau and Weinzierl \cite{lindau2012assessing} proposed a yes/no paradigm to assess the plausibility of virtual versus real sources, comparing non-individual HRTFs for five loudspeaker positions with their real loudspeaker counterparts. They found that plausibility can be achieved in dynamic binaural auralisations and that it is less sensitive to spectral colouration deviations than authenticity.

Authenticity requires a direct comparison between a virtual and a real sound source, with the highest authenticity possible achieved only when the two are indistinguishable. Plausibility, on the other hand, does not involve such a direct comparison; it is an absolute judgment of whether the virtual sound source is perceived as real or virtual. When assessing the performance of an AAR module, plausibility offers a less strict criterion but is arguably more relevant, as the goal of AAR is to ensure that the virtual sound source blends seamlessly into the surrounding sound scene, rather than being perceptually identical to a specific real-world reference. 

This study explores whether the use of individual HRTFs is essential for the localisation and realism of virtual sound sources depending on the presence or absence of head movements. The protocol simulates a simple AAR scenario where virtual sound sources are associated with real visual objects (loudspeakers). Stimuli are either generated by the real source, or binaurally through a pair of hear-through headphones. Participants are asked to localise the target sources and rate them in terms of realism and externalisation. Speech recordings are used in order to provide a reference for the realism ratings required in this study, as normal-hearing individuals have a consistent understanding of how spoken voices sound in natural environments. The first hypothesis is that the benefit of head movements would surpass the advantages of HRTF individualisation on localisation performance. Conversely, it is also hypothesised that the freedom to make head movements might decrease the impression of realism when compared to real sound sources as it becomes more obvious that the sound is emitted from headphones, rendering the benefits of HRTF individualisation for realism negligible.

In the manuscript, a first section presents the experimental paradigm of the study, how and in which context auditory stimuli were rendered, and how data was collected from participants. A second section presents the results and the different statistical analyses. Finally, the results of the perceptual evaluations are discussed in relation to the initial hypotheses.


\section{Methods}\label{sec:methods}

\subsection{Material and listening environment}

All stimuli were generated using anechoic speech recordings from the TSP speech database \cite{kabal2002tsp}. Twenty-five 3-second-long recordings, each corresponding to a different talker, were used. At each trial, a recording was chosen randomly within the dataset, and could not be used for the next trial. Switching to a different speech recording for each trial avoided the presentation of sequential stimuli, which could reveal obvious spectral differences.

\begin{figure}[H]
    \centering
    \includegraphics[width=0.9\textwidth]{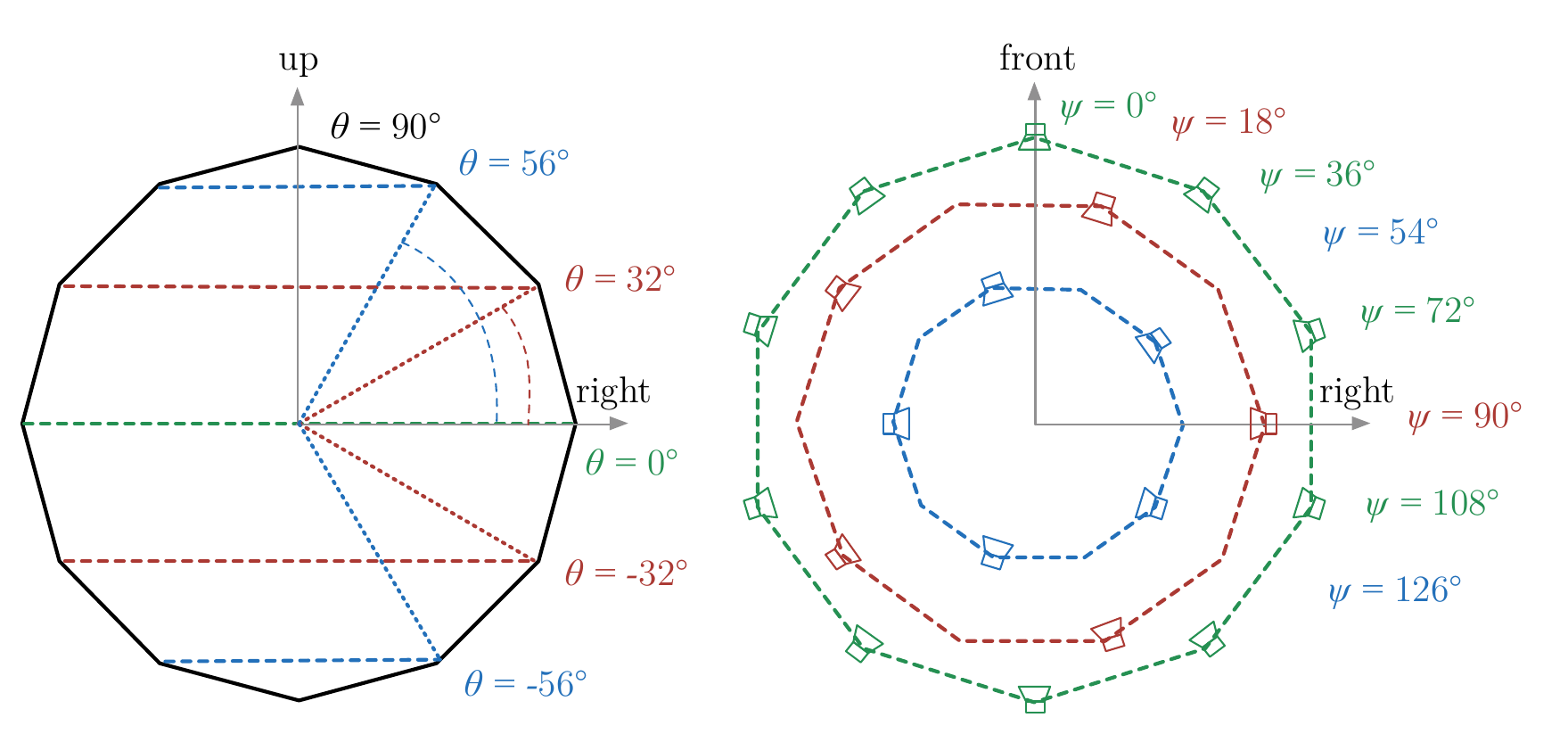}
    \caption{The spherical loudspeaker array layout utilised for the experiment. The participant was seated at the centre of the sphere. The array consists of 30 loudspeakers organised across 5 distinct planes at varying elevations. The two planes shown in blue are positioned at an elevation $\theta = [-56^{\circ}, 56^{\circ}]$, and comprise 5 loudspeakers each equally spaced at intervals of $72^{\circ}$ of azimuth. The planes depicted in red, with an elevation of $\theta = [-32^{\circ}, 32^{\circ}]$, also comprise 5 loudspeakers each, equally spaced. The horizontal plane, shown in green at $\theta = 0^{\circ}$, comprises 10 loudspeakers equally spaced at $36^{\circ}$ of azimuth.}
    \label{fig:loudspeakerarray}
\end{figure}

The experiment was carried out in a semi-anechoic room (RT60(1kHz)=0.1s, 22dBA ambient noise level). This is the same room used for the HRTF measurements \cite{engel2023sonicom} referenced later in the manuscript. However, the reverberation time deviates slightly from the values reported in \cite{engel2023sonicom} due to changes in the room’s configuration; more specifically, for the present experiment the space was furnished with a 30-loudspeaker spherical array (see Figure \ref{fig:loudspeakerarray}), which resulted in alterations of the acoustic properties of the environment.  Each loudspeaker in the array was labelled in terms of its elevation (high, mid-high, middle, mid-low and low) and azimuth (1 to 10 for the ``middle" circle, 1 to 5 for the others) for localisation reporting.

\subsection{Stimuli rendering}

``Real" stimuli were presented via the loudspeakers (\textit{Genelec 8010A}), calibrated so the signals were delivered at an average loudness of 68dBA at the center of the loudspeaker sphere. A custom \textit{Max/MSP} patch managed stimulus delivery via an \textit{Antelope Audio Orion 32} audio interface.

``Virtual" stimuli were presented through a pair of \textit{AKG K1000} open-ear headphones powered by a \textit{SMSL AO200} amplifier. Using again \textit{Max/MSP}, together with the \emph{3DTI Toolkit} \cite{cuevas20193d} spatialiser, and an \textit{RME Madiface} audio interface, loudspeakers were virtualised through three different types of binaural rendering. To take advantage of the high spatial resolution of the HRTFs without being limited by the spatial restraints of the Ambisonics-based reverb employed here (fourth-order), the direct path component of the auditory stimuli was processed separately from the reflected ones. 

To produce the direct sound, the anechoic recordings were convolved with a Head-Related Impulse Response (HRIR) dataset; specifically, the individual HRTF of each participant was employed, as well as the dummy head HRTF (KEMAR) measured with the same measurement setup \cite{engel2023sonicom}.

While the HRTF from \cite{engel2023sonicom} were available also in the un-windowed version (i.e. with room reflections), it was decided to use the windowed HRTFs (i.e. with room reflections windowed out) because of the following reasons:
\begin{itemize}
    \item Interpolation of HRTFs with room reflections is potentially problematic, especially due to the different timings of the echoes from the different HRTF positions.
    \item The room configuration during the experiment differed from that used during the HRTF measurements, specifically due to the presence of the 30-loudspeaker array.
\end{itemize}

Reverberation was therefore rendered by convolving the anechoic recordings with fourth-order Ambisonics Room Impulse Responses (RIRs), measured separately for each of the 30 possible sound source positions.

Measurements were performed with an \textit{EigenMike EM32} spherical microphone array at the position of the participant corresponding to the center of the sphere. These RIRs had the direct sound component removed by substituting the initial part, up to $4.5$ ms after the onset, with zeroes, ensuring that the delay between direct and reflected components is preserved at the playback stage. This threshold was chosen in order to include all first order reflections in the reverb rendering.

The convolution of the reverberant part resulted in an HOA signal, which was then decoded into a binaural format using the so-called ``virtual loudspeakers" paradigm \cite{engel2021perceptual}. An array of $25$ virtual loudspeakers, positioned on the vertices of a dodecahedron, was employed. Based on the position of each virtual loudspeaker on the chosen virtual grid, the associated HRIRs for both ears were identified. Each of them was convolved with that specific loudspeaker feed, and the convolution products were then summed (separately for the left and right channels), generating the binaural signal for each ear.

The levels of the direct sound and reverberation were determined so that the direct-to-reverberant ratio (DRR) was kept equal to the real measurements made in the room, for each source. The use of the \emph{3DTI Toolkit} allowed for the sound scene to be modified dynamically using real-time head tracking information, which resulted in a change in the listener's rotation within the simulated environment, and a consequent change in the relative source positions for both the direct sound and the reverberation. (i.e. virtual loudspeakers).

The \textit{HTC Vive} tracker used for head tracking has a reported (by the manufacturer) latency of 10 ms, to which the audio processing delay from a 256-sample buffer at 48 kHz (5.33 ms) was added, resulting in an estimated total of 15.3 ms. This is well below the 30 ms detectability threshold for head-tracker latency reported by Brungart et al. \cite{brungart2005detectability}.

In addition to the individual and KEMAR HRTF conditions described above, which represent a ``faithful" reproduction of the real scene, a condition with an exaggerated reverberation was generated by raising its level by 12dB. This was produced only with the KEMAR HRTF, and was used in order to provide an ``anchor" with low perceived realism and high externalisation. It was expected that, because of the semi-anechoic nature of the room, other rendering methods would induce relatively low externalisation ratings.

In order to match the response of the various rendering and processing chains, a series of equalisation filters was employed. The loudspeaker rendering was used as a reference, and the direct and reverberant components of the signals were processed independently. 
First, a KEMAR dummy head wearing the \textit{AKG K1000} headphones used for the test was placed in the position of the participant. The impulse response of the front loudspeaker was measured, and direct and reflected components were separated. The same was done using the headphones as playback device, and spatializing the source in a frontal position employing the KEMAR HRTF for the direct path and the Spatial Room Impulse Response recorded with the \textit{Eigenmike EM32} for the reflected one. The approach detailed in \cite{engel2019effect} was then employed to match the headphone response to the loudspeaker one, independently for the direct and reflected paths, generating two 1024 taps Finite Impulse Response (FIR) filters. These were then applied to the processing of the headphones stimuli in \textit{MAX/MSP}. Finally, both filters were fine-tuned manually through a series of informal listening tests, in order to carefully reduce any potential spectral difference.

\subsection{Conditions}

Four different sound renderings were evaluated:
\begin{itemize}
    \item Real sound sources produced by loudspeakers
    \item Binaural sources rendered with an individual HRTF and same DRR as real sources
    \item Binaural sources rendered with the KEMAR HRTF and the same DRR as real sources
    \item Binaural sources rendered with the KEMAR HRTF and an exaggerated reverberation    
\end{itemize}

Thirty different positions (corresponding to the 30 loudspeakers of the array) were evaluated once per rendering method, resulting in a total of 120 stimuli per session. All binaurally rendered stimuli were spatialised at the location of the real loudspeakers. For each of the stimuli, a random anechoic recording from the database of speech stimuli was selected. All stimuli were generated in real-time.

Two different sessions were run, the second spaced a minimum of two weeks after the first. In the first, participants could use head movements during stimulus presentation. In the second, the head movements were minimized using a chin rest. The order of the head-tracked and fixed-head sessions was counterbalanced, with half (5) of the participants starting with each condition to control for learning effects.

\subsection{Procedure and collection method}

\begin{figure}[H]
    \centering
    \includegraphics[width=0.7\textwidth]{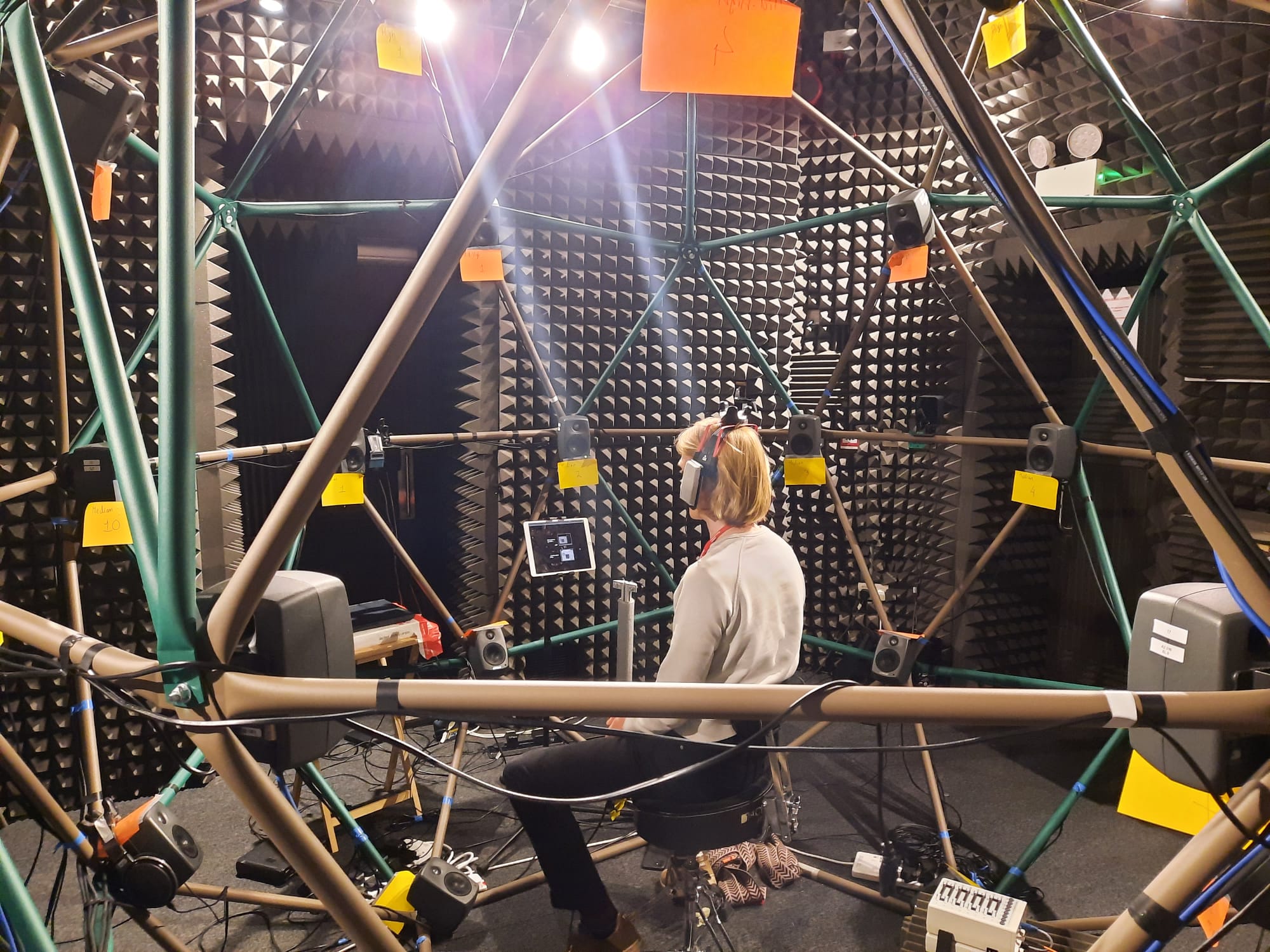}
    \caption{Experimental setup used for both sessions of the experiment. The participant's head is located at the center of the spherical loudspeaker array, and a touch tablet is positioned in front of the participant to report responses.}
    \label{fig:setup}
\end{figure}

Participants were introduced to the experiment and signed an informed consent form. They were then seated on a rotating chair and introduced to the graphical interface presented on a tablet in front of them (cf.  Figure \ref{fig:setup}). Participants rated the attributes of different stimuli on a visual scale with the exception of localisation, for which they used a simple graph to indicate the perceived position of the sound source. To report localisation, participants were first encouraged to visually inspect the surrounding loudspeakers, each labelled with its corresponding number and elevation, after listening to the stimulus and before entering their response on the tablet. The response format was discrete, requiring the selection of one loudspeaker position from the set, making the method closer to an egocentric head-pointing approach \cite{bahu2016comparison} that has been shown to be precise when visual cues are present \cite{tabry2013influence}. In this setup, the touch tablet served only to record the participant’s choice after visually identifying the source position in the loudspeaker sphere.

The other attributes and associated rating range were:

\begin{itemize}
    \item Realism, on a slider scale going from 0 to 100. It was explained that the value 0 corresponds to complete confidence that the sound is coming from headphones and 100 to complete confidence it is coming from the loudspeakers.
    \item Externalisation, on a slider scale from 0 to 5. It was explained that 0 corresponds to the perception of a sound coming from the inside of the head and 5 from outside the head (i.e. indicating ``natural" externalised listening).
\end{itemize}

Realism was rated on a continuous 0 to 100 scale to capture participants’ confidence in whether a stimulus sounded “real” or “virtual” with fine granularity, and to reduce bias toward categorical choices, which is relevant as some participants were familiar with spatial audio and binaural rendering. Unlike stricter, binary “authenticity” scales often used in binaural research \cite{brinkmann2017authenticity, starz2022perceptual}, the realism scale used here was intentionally closer to a "plausibility" rating as it can be evaluated in \cite{starz2025comparison}. The aim was not to determine whether the reproduction was indistinguishable from reality, but rather whether it could be perceived as a real sound source. externalisation, being more of a categorical percept \cite{best2020sound}, was rated on a 1 to 5 scale. 

During the session with head movements, participants were told to use the strategy they thought was the best to localise the stimulus. Before its presentation, participants had to face the loudspeaker at an azimuth angle of $0^{\circ}$ on the horizontal plane, and could start to move their head after the stimulus started. In the session without head movements, they had to keep their head on a chin rest oriented towards the same centre loudspeaker.

After being informed about the approximate duration of the experiment (40 minutes), participants engaged in a training session consisting of 20 stimuli using all the four renderings (3 binaural renderings + loudspeakers) and one position from each of the elevation planes (5). The goal of the training was to familiarze the participant with the reporting method and ensure that the procedure was understood. Feedback about the location and the rendering method (``loudspeaker" or "headphones") was given to the participant after each stimulus during the training session.

\begin{figure}[H]
    \centering
    \includegraphics[width=0.7\textwidth]{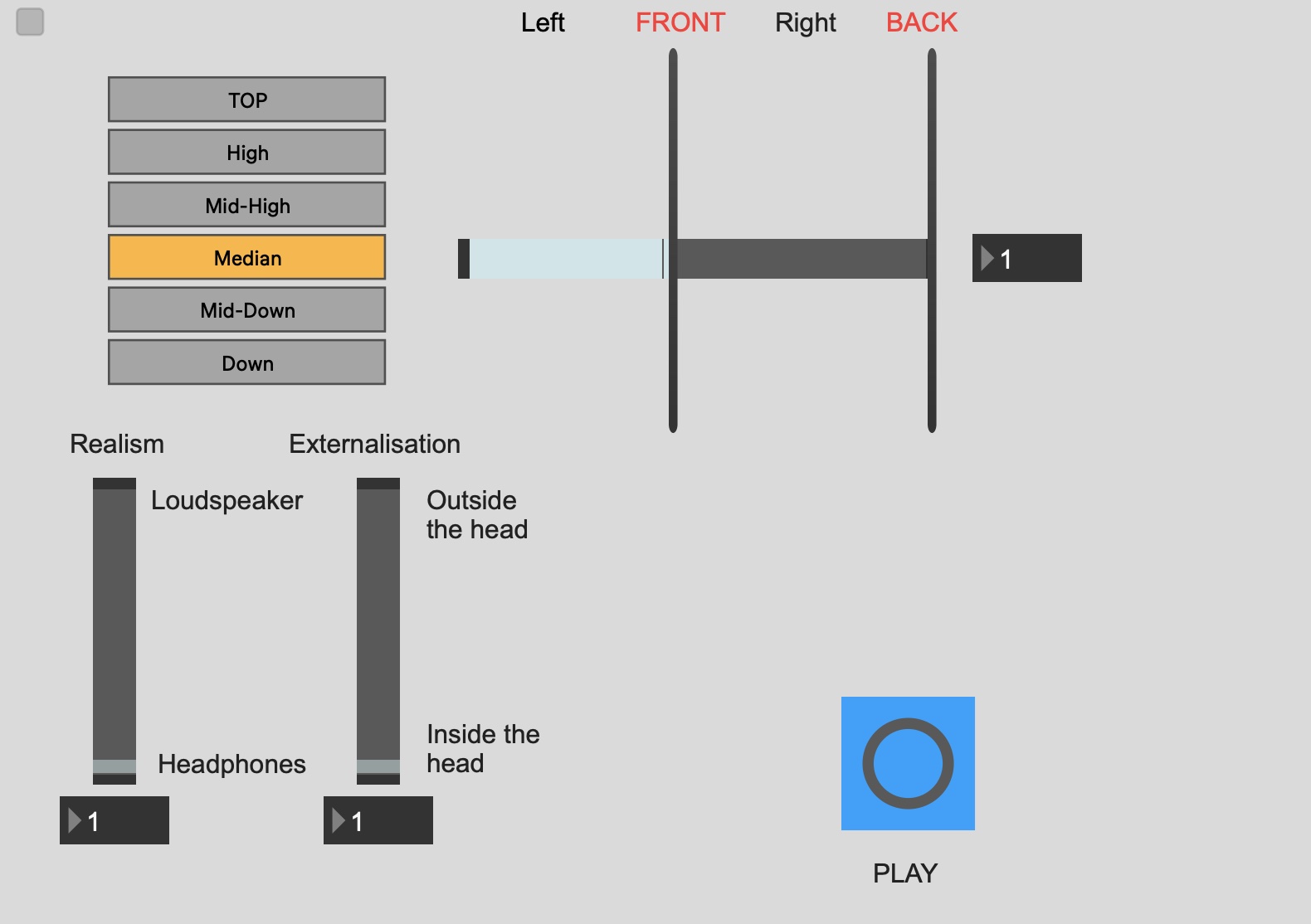}
    \caption{Interface displayed on the touch tablet used by the participant to report localisation (designating one loudspeaker location on the sphere), realism (from 1 to 100) and externalisation from (1 to 5).}
    \label{fig:interface}
\end{figure}

After the training, the 120-trial experiment started. The order of the stimuli, their location and the rendering method were randomised. Participants could trigger the stimulus playback only once, with each stimulus being played for 3 seconds. Participants were then asked to report on the perceived location and on the different attributes (see interface in Figure \ref{fig:interface}). The specific duration of 3 seconds was chosen in order to allow enough time for the listener to perform head movements, and at the same time not to be too long (making it harder for the participant to remain still) for the static condition. 

\subsection{Localisation performance metrics}\label{subsec:metrics}

In order to evaluate localisation performance, usual metrics for assessing binaural localisation  were first considered, following the guidelines reported by Poirier-Quinot et al. \cite{poirier2022hrtf}. However, as the localisation reports are on discrete positions, the standard continuous metrics (polar/lateral precision and accuracy) were ruled out.

A new set of metrics has therefore been chosen, the first being the confusion rate calculated with the Great Circle Error (GCE). The GCE is defined as the minimum arc (in degrees) between the reported location and the true target location. Although the GCE is a continuous metric, it is the only evaluation that allows the classification of confusion errors. While the discrete nature of the responses might introduce some errors in the classification, it is very unlikely for this to happen for front-back confusions. Therefore, the analysis of the results will focus particularly on these types of errors when examining the confusion rates. This metric provides an intuitive way to assess the local localisation accuracy as the spherical distance between the responses and the target. Given $xyz_{target}$ and $xyz_{response}$ as the vectors in Cartesian coordinates of the target and response positions respectively, the GCE is defined as:
\begin{equation*}
    GCE = arctan(  \frac{\left \| xyz_{target} \times  xyz_{response} \right \| }{xyz_{target} \cdot xyz_{response} })
\end{equation*}

From the evaluation of the GCE responses, 4 categories can be created: Precision errors occur for responses near the target with a GCE $< 45^\circ$. Front-back confusions occur for responses within a GCE $< 45^\circ$ around the symmetrical target position regarding the frontal plane. In-cone errors occur for responses with a lateral angle within $45^\circ$ of that of the target, that are not already classified as either precision or front-back confusions. Off-cone errors are the rest of the responses not falling in the above categories.

Considering that the standard continuous metrics were ruled out due to the discrete nature of the reported positions, the second metric used is the rate of responses that reported the correct loudspeaker location. 

To further investigate the impact of the rendering and the presence of head movements on localisation performance, spatial decomposition of the source positions based on lateralisation was considered. The 30 different sources positions were divided into 15 ``unlateralised" positions in the front and back quadrants (with an azimuth within $[315^\circ;45^\circ]\cup[135^\circ;225^\circ]$) and 15 ``lateralised" positions in the lateral quadrants (with an azimuth within $[45^\circ;135^\circ]\cup[225^\circ;315^\circ]$).

\subsection{Participants}

The experiment involved 10 participants (3 females, 7 males, mean age: 32). All of them were experts in spatial audio, and had their individual HRTFs acoustically measured with the same setup \cite{engel2023sonicom}. All of them signed an informed consent before participating in the experiment and reported no hearing impairments. All were asked to complete a questionnaire in which they reported their gender, age, and previous experience with spatial audio. 
The study and methods followed the tenets of the Declaration of Helsinki. Approval for the experiment design was given by the Science, Engineering, and Technology Research Ethics Committee (SETREC, reference: 21IC6923) at Imperial College London. Data acquisition followed the guidelines of the general data protection regulation (GDPR).

\section{Results}\label{sec2}

The first analysis focuses on the localisation performance results of the 10 participants, while the second looks at perceived realism and externalisation. Both sections include results for conditions with and without head movements.

The localisation performance analysis is based on one Repeated Measures Analysis of Variance (RM-ANOVA) per session (with and without head movements) with rendering method and lateralisation as  within-subject factors. For realism and externalisation, one RM-ANOVA per data type collected and session was performed, with rendering method as within-subject factor. 

The mean ratings (over all 30 source locations) per participant for each rendering method were treated as the dependent variable. A significance value of $\alpha = 0.05$ was used. Initial focus was placed on the identification of the outliers. The detection was performed separately for each condition. Data points were considered outliers if they fell more than 1.5 times the interquartile range (IQR) below the first quartile or above the third quartile within that condition. In total, five data points were identified as outliers across all variables (160) studied. Given the small number relative to the total dataset and that they are not relied to a single participant, these points were retained in the statistical analyses. This approach ensured that the results reflected the full set of observations. To maintain transparency, data points identified as outliers are highlighted in the figures of this section.


The normality of the variables was evaluated using the Shapiro-Wilk test. Sphericity was examined with Mauchly’s test. If Mauchly’s test indicated a violation of sphericity ($p < 0.05$), corrections were applied based on the Greenhouse–Geisser epsilon: the Huynh–Feldt correction was used if $\epsilon > 0.75$, and the Greenhouse–Geisser correction was used if $\epsilon \leq 0.75$, in order to control for Type I errors.


\subsection{Localisation performance}

\begin{figure}[H]
    \centering
    \includegraphics[width=0.8\linewidth]{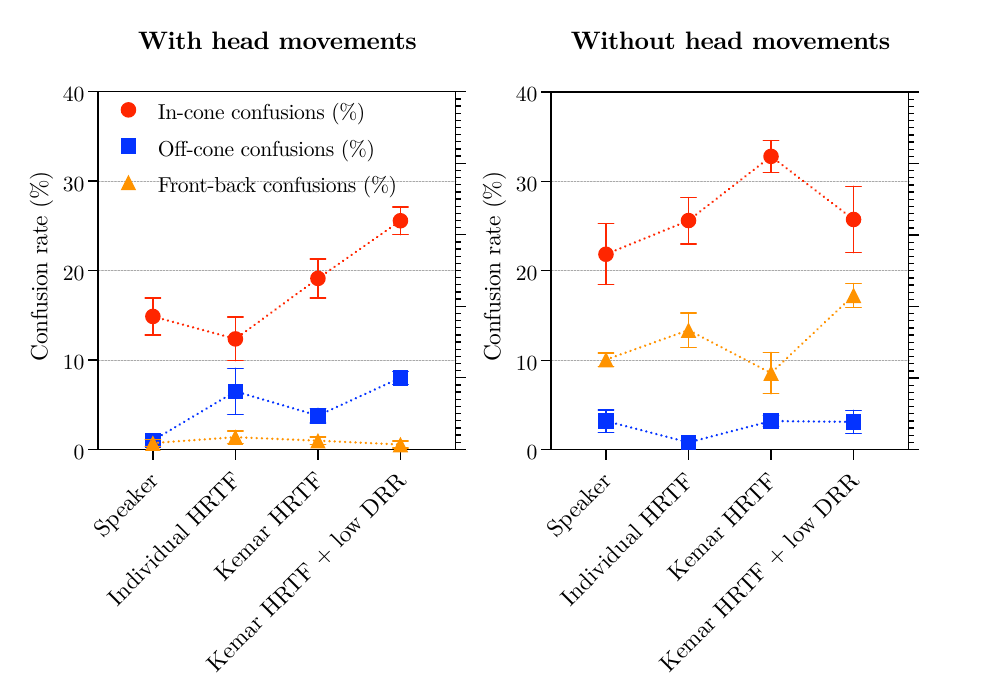}
    \caption{Mean rate of response confusions for each session with head movements (left) and without head movements (right). Error bars indicate the 95 percent confidence intervals (CIs) for each type of confusion calculated over all the responses collected for a rendering method.}
    \label{fig:confusion}
\end{figure}

The metrics introduced in Section \ref{subsec:metrics} obtained from the 10 participants' reports during both sessions are presented here. The GCE on all localisation reports was calculated relative to the real location of the targets. Thus, the localisation reports could be classified into four types of confusion errors (precision, in-cone, off-cone and front-back). The classification results are illustrated in Figure \ref{fig:confusion}. Descriptive analysis shows that stimuli rendered with the loudspeakers are associated with less confusion. Moreover, the presence of head movements drastically reduces confusion rates, especially for front-back confusions. 

\begin{figure}[H]
\centering

\begin{subfigure}[b]{1\textwidth}
   \includegraphics[width=1\linewidth]{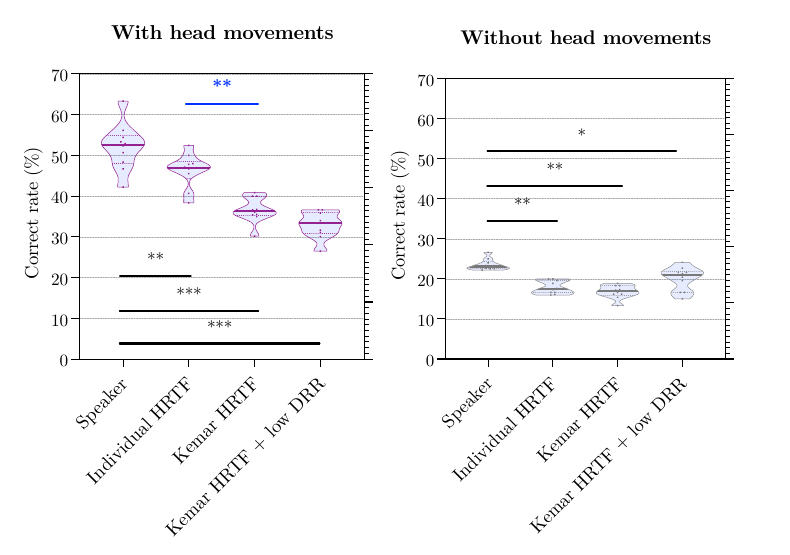}
\end{subfigure}

\caption{Mean rate of correct responses per participant for each session with head movements (left) and without (right) represented by violin plots, displaying the probability density of the data and the overall mean value (horizontal line). The symbols (*) represent p-values $< 0.05$, (**) represent p-values $< 0.01$, and (***) represent p-values $< 0.001$, indicating the respective levels of statistical significance of the post-hoc tests.}
\label{fig:correctrates}
\end{figure}

In order to further investigate the influence of rendering on localisation performance, the rates of correct localisation response were calculated. The correct rates as a function of the rendering method are presented in Figure \ref{fig:correctrates} and as a function of the lateralisation in Figure \ref{fig:correctratezone}. Descriptive analysis shows that loudspeaker rendering results in the best localisation performance, and that head movements drastically improve localisation.

To study the statistical significance of the rendering method on the rate of correct response in the session with and the one without head movements, two separate RM-ANOVAs (one for each session) were conducted with the rendering method and the lateralisation of the source as factors. The variables involved passed Mauchly's test for sphericity ($p > 0.05$), and Shapiro-Wilk's test for normality (for all levels $p > 0.05$). The results of these statistical analyses are presented in Table \ref{tab:correctrate}. A significant effect of rendering method and of the interaction were found. However, no significant main effect of the lateralisation was found.


Post-hoc tests with Bonferroni-Holm correction were conducted to compare the localisation performance for each rendering method (4). The p-value of these t-tests was adjusted for a family of 6 estimates. The results are reported in Figure \ref{fig:correctrates}.

For both sessions, the results show that the errors associated with loudspeakers are significantly lower than those from all other renderings. In contrast, the errors for the rendering using the KEMAR HRTF with exaggerated reverb are significantly higher. Differences due to the presence or absence of head movements are evident in the post-hoc tests comparing metrics for individual and KEMAR HRTF renderings. With head movements, rendering with the individual HRTF results in significantly fewer errors than with the KEMAR HRTF. Without head movements, post-hoc tests do not reveal significant differences between these two rendering methods.

\begin{table}[ht]

\begin{subtable}{0.7\textwidth}
\sisetup{table-format=-1.2}   
\centering
   \begin{tabular}{lrrrrr}
			\toprule
\textit{\textbf{}}			  & \textit{\textbf{df}} & \textit{\textbf{Mean Square}} & \textit{\textbf{F}} & \textit{\textbf{p-value}} & \textit{\textbf{$\eta^{2}_{p}$}}  \\
			\cmidrule[0.4pt]{1-6}
			Rendering & $3$ & 1,679.012 & $87.066$ & $<$ .001 & $0.906$  \\
			
			Lateralisation & $1$ & $24.741$ & $0.910$ & $.365$ & $0.092$  \\
			
			Rendering * Lateralisation & $3$ & $381.157$ & $46.964$ & $<$ .001 & $0.839$  \\
			
			\bottomrule
		\end{tabular}
   \caption{With head movements}\label{tab:correctrateHM}
\end{subtable}

\begin{subtable}{0.7\textwidth}
\sisetup{table-format=-1.2}   
\centering
   \begin{tabular}{lrrrrr}
			\toprule
			 &  \textit{\textbf{df}} & \textit{\textbf{Mean Square}} & \textit{\textbf{F}} & \textit{\textbf{p-value}} & \textit{\textbf{$\eta^{2}_{p}$}}  \\
			\cmidrule[0.4pt]{1-6}
			Rendering  & $3$ & $187.978$ & $48.889$ & $<$ .001 & $0.845$  \\
			Lateralisation & $1$ & $47.104$ & $3.363$ & $.100$ & $0.272$  \\
			Rendering * Lateralisation & $3$ & $40.045$ & $6.724$ & $.002$ & $0.428$  \\
			\bottomrule
		\end{tabular}
   \caption{Without head movements}\label{tab:correctratenoHM}
\end{subtable}

\caption{Results from four RM-ANOVAs with the rendering method and lateralisation considered as within-subject effects, for the rate of correct response of each participant (10 participants) in the session with head movements (a) and without head movements (b).} \label{tab:correctrate}
\end{table}

\subsection{Localisation performance per zone of appearance}

\begin{figure}[H]
    \centering
    \includegraphics[width=0.8\linewidth]{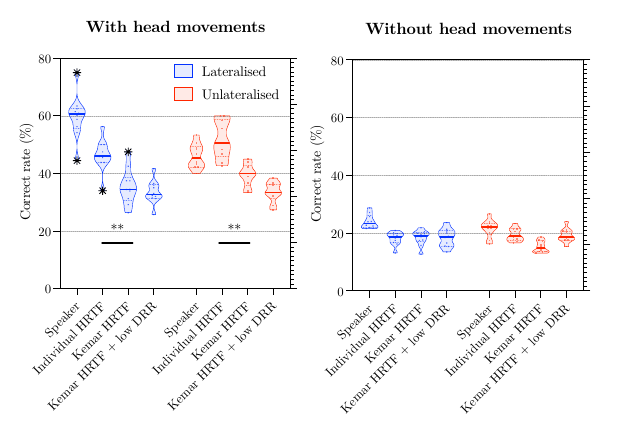}
    \caption{Mean rate of response for "lateralised" and "unlateralised" sources for each session with head movements (left) and without head movements (right)  represented by violin plots, displaying the probability density of the data and the overall mean value (horizontal line). Outliers detected with the 1.5$\times$IQR rule are highlighted in black. The symbols (*) represent p-values $< 0.05$, (**) represent p-values $< 0.01$, and (***) represent p-values $< 0.001$, indicating the respective levels of statistical significance of the post-hoc tests.}
    \label{fig:correctratezone}
\end{figure}

To further investigate the impact of the rendering and the presence of head movements on localisation performance, spatial decomposition of the targets based on the lateralisation was considered and integrated into the RM-ANOVAs which results are presented in Table \ref{tab:correctrate}. The spatially decomposed localisation performance of participants is illustrated in Figure \ref{fig:correctratezone}. Therefore, further post hoc tests with Bonferroni-Holm correction were performed between data for each rendering method and lateralisation condition ($4\times2$). The p-value of these t-tests was adjusted for a family of 28 estimates. The difference between individual HRTF and KEMAR HRTF in each lateralisation condition are reported in Figure \ref{fig:correctratezone}.

With head movements, the results reveal that individual HRTF leads to significantly better localisation performance in both lateralisation conditions. Without head movements, there are no significant differences in localisation performance between the two rendering methods, in both lateralisation condition.

\subsection{Realism and externalisation}

\begin{figure}[H]
\centering
\begin{subfigure}[b]{1\textwidth}
   \includegraphics[width=1\linewidth]{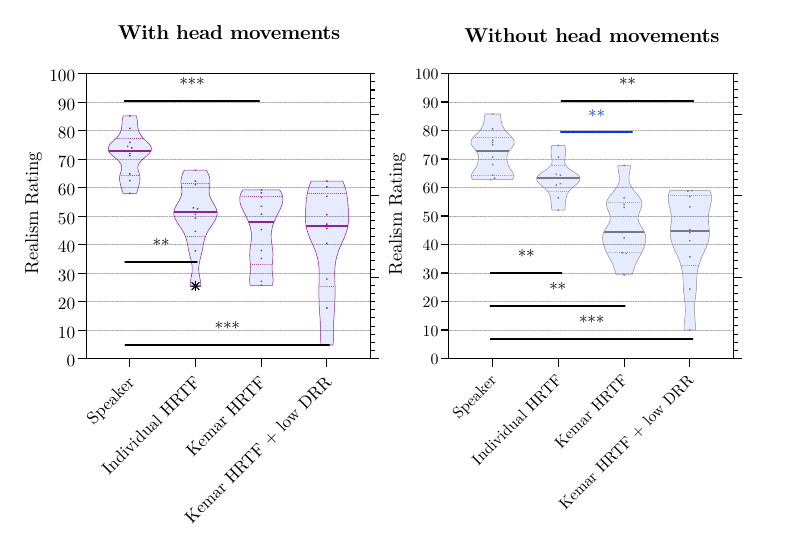}
\end{subfigure}

\begin{subfigure}[b]{1\textwidth}
   \includegraphics[width=1\linewidth]{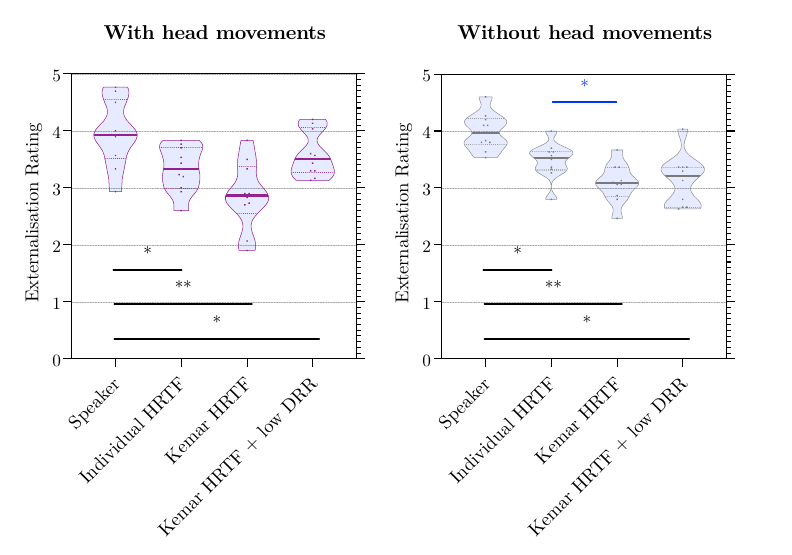}
\end{subfigure}

\caption{Mean realism (above) and externalisation (below) ratings per participant for each session with (left) and without (right) head movements represented by violin plots. It shows the probability density of the data and the overall mean value (horizontal line). Outliers detected with the 1.5$\times$IQR rule are highlighted in black. The symbols (*) represent p-values $< 0.05$, (**) represent p-values $< 0.01$, and (***) represent p-values $< 0.001$, indicating the respective levels of statistical significance of the post-hoc tests.}
\label{fig:realismexternalisation}
\end{figure}

The mean ratings per participant associated with each rendering method in terms of realism and externalisation are illustrated in Figure \ref{fig:realismexternalisation}. Descriptive analysis indicates that stimuli rendered with loudspeakers are associated with higher realism and externalisation ratings, compared to binaural renderings.

To study the influence of the rendering method on realism and externalisation, one RM-ANOVA for each of the mean ratings was conducted. The dependent variable was the rendering method, and the results are presented in Table \ref{tab:realismexternalisation}. The concerned variables passed the Mauchly's test for sphericity ($p > 0.05$), and Shapiro-Wilk for normality (for all levels $p > 0.05$).

These analyses reveals a significant main effect of the rendering method on the externalisation and the perceived realism. To further investigate specific differences, post-hoc tests with Bonferroni-Holm correction were conducted on the ratings associated with each rendering methods. The p-value of these t-tests were adjusted for a familily of 6 estimates. These results are illustrated in Figure \ref{fig:realismexternalisation}. The post-hoc tests confirm that the rendering through loudspeakers consistently result in higher realism and externalisation ratings. Head movements induced differences when comparing the results of the renderings between individual and KEMAR HRTF. In contrast to the localisation performance metrics, when head movements are used, no significant differences are found in externalisation and realism ratings between individual and KEMAR HRTFs. Without head movements, individual HRTF leads to significantly higher realism ($p=0.004$) and externalisation ($p=0.14$) ratings than KEMAR HRTF.

\begin{table}[h]

\begin{subtable}{0.7\textwidth}
\sisetup{table-format=-1.2}   
\centering
\begin{tabular}{lccccc}
\hline
\textit{\textbf{Ratings}} &
  \textit{\textbf{Sphericity correction}} &
  \textit{\textbf{df}} &
  \textit{\textbf{F}} &
  \textit{\textbf{p-value}} &
  \textit{\textbf{$\eta$$^{2}$}} \\ \hline
Realism &
  None &
  3 &
  16.8 &
  $<$ 0.001 &
  0.707 \\
Externalisation &
  None &
  3 &
  4.6 &
  0.02 &
  0.375 \\ \hline
\end{tabular}
   \caption{With head movements}\label{tab:realismextern_HM}
\end{subtable}

\begin{subtable}{0.7\textwidth}
\sisetup{table-format=-1.2}   
\centering
  \begin{tabular}{lccccc}
\hline
\textit{\textbf{Ratings}} &
  \textit{\textbf{Sphericity correction}} &
  \textit{\textbf{df}} &
  \textit{\textbf{F}} &
  \textit{\textbf{p-value}} &
  \textit{\textbf{$\eta$$^{2}$}} \\ \hline
Realism &
  None &
  3 &
  9.6 &
  $<$ 0.001 &
  0.579 \\
Externalisation &
  None &
  3 &
  13.2 &
  $<$ 0.001 &
  0.654 \\ \hline
\end{tabular}
   \caption{Without head movements}\label{tab:realismextern_noHM}
\end{subtable}

\caption{Results from four RM-ANOVAs with the rendering method considered as within subject effect, for the mean realism and externalisation rating.  The dependent variable considered was the averaged rating for each participant (10 participants) in the session with head movements (a) and without head movements (b).} \label{tab:realismexternalisation}
\end{table}

\section{Discussion}

This study focuses on the influence of head movements and the use of individual or non-individual HRTFs in a real/virtual localisation task. The results show that individual HRTFs result in lower localisation errors compared to the KEMAR HRTF when head movements are available. However, when considering externalisation and realism, the results are inverted, with the impact of individual HRTFs observed only in the static condition.

\subsection{Localisation performance}

In general, results show that the presence of head movements is strongly beneficial for all rendering methods. For all binaural renderings, it removed almost all front-back confusions from the reports (see Figure \ref{fig:confusion}). This result can be comparable to the study by Oberem et al. \cite{oberem2020experiments}, who found that dynamic listening substantially increases localisation accuracy, improves externalisation, and reduces front-back confusions, while also reducing the observable advantage of individualised over non-individualised HRTFs.

Results from localisation performance with loudspeaker rendering indicate the difficulty of the task, with participants averaging a 53\% correct rate with head movements and 23\% without, despite all participants being experts in spatial audio. This difficulty could first be caused by the spatial density of the different sources. In the study conducted by Gaveau et al. \cite{gaveau2022benefits} where participants had to localise real sound sources with head movements, the mean angular errors were on average 21.6$^\circ$. A similar study by Wightman et al. \cite{wightman1999resolution} found an average error of 16.3$^\circ$. Both these values are close to the minimum angle between consecutive loudspeakers in the array used in this study which is 18$^\circ$.     
The length of the stimuli (3 seconds only played once) could also be a factor. A study by Thurlow et al. \cite{thurlow1970effect} compared different stimulus duration for the localisation performance of real sound sources in a dynamic listening situation. They demonstrated than a 2-second stimulus was close to the minimum to gain for the benefit of the head movements (time for participants to turn their head towards the source) but that 5-second stimuli led to better performances.

A statistical analysis conducted on the rate of correct localisation responses showed that the individualisation of HRTFs led to better localisation performance in the session with head movements only. While it is generally known that HRTF individualisation is beneficial for localisation performance \cite{wenzel1993localization}, Begault et al. \cite{begault2001direct} showed that its benefit is generally not significant when the head movements are tracked. However, Begault et al. study only focused on sources present in the horizontal plane at the listener's ear level. In this study, the sound sources were surrounding the participant at different elevation planes. This can be explained by the fact that sources on the horizontal plane are generally easier to localise than elevated ones \cite{makous1990two,blauert1997spatial}; as such, the benefits of head movements may not be apparent when averaging across all source locations on a sphere. Therefore, in this study, the observed improvement in localisation performance with head movements could be due to sources outside the horizontal plane.

Moreover, when decomposing the results as a function of the lateralisation of the sources' location (see Figure \ref{fig:correctratezone}), the localisation of both ``lateralised" and ``unlateralised" benefited from the individualisation of HRTF in the session with head movements. This can be explained by the nature of the head movements used by the participants. Similar studies have shown that when localising sound sources using head movements, listeners tend to turn their heads on the horizontal plane towards the azimuth of the source  \cite{thurlow1970effect, gaveau2022benefits}. Therefore, with head movements the lateralisation of the sources becomes a less critical factor on localisation performance, as both types of sources rely on the same binaural and visual cues.



Oberem et al. \cite{oberem2020experiments} reported a non-significant trend toward improved localisation for front and back sources only, when individualised HRTFs were employed in an experiment without head movements. In their study, the localisation accuracy measures used (unsigned vertical and horizontal error, in-head localisation rate, and front–back reversal rate) remained very low for “lateralised” sources, which may have compressed observable differences between conditions. In the present study, this observation could not be confirmed with the results of the session without head movements. In the session without head movements, the benefit of HRTFs individualisation on localisation was not significantly affected by source lateralisation.

\subsection{Realism and externalisation}


First, it should be noted that binaural rendering with faithful reproduction of the reverberation produced a generally high level of realism. All participants, who were experts in spatial audio, reported that it was sometimes very difficult to determine whether the sound source was produced by a loudspeaker or by the headphones. This was facilitated by the pseudo-anechoic nature of the environment, which did not require the simulation and matching of complex acoustic conditions in the virtual rendering. This led to participants rating loudspeaker stimuli with scores below 100, indicating less than complete confidence that the sound was from a real source. This cautious behaviour was due to the desire to avoid being misled by stimuli originating from the headphones.
A strong correlation was observed between externalisation and realism ratings, with post-hoc tests revealing consistent differences between the rendering methods for both measures. The only exception was the binaural rendering with exaggerated reverb, which was included in the protocol to decorrelate such characteristics. While it is generally assumed that externalisation is a prerequisite to realism \cite{best2020sound}, lower DRRs increase externalisation but reduce realism due to a stronger room divergence effect \cite{best2020sound, werner2016summary}. Therefore, it was expected that the rendering method with a lower DRR would induce higher externalisation but poorer realism.

Comparisons between conditions with and without head movements must be made cautiously as the experiments were conducted at different times. Nonetheless, no significant improvement in externalisation and realism ratings was observed with head movements. This result shows that our current realism evaluation, based on plausibility, can yield different outcomes compared to stricter authenticity assessments that rely on direct comparisons. For instance, Brinkmann et al. \cite{brinkmann2017authenticity} employed a two-alternative forced-choice (2AFC) test in which participants compared a real and a virtual stimulus in each trial, and found that head movements led to a decrease in perceived authenticity. The present study's result aligns with findings by Begault et al. \cite{begault2001direct}, who also studied the effect in anechoic conditions with a similar stimulus length (2-3 seconds). In contrast, Hendrickx et al. \cite{hendrickx2017influence} found improvements in a reverberated condition with longer stimuli (8 seconds) and suggested that head movements need to be sufficiently large to have a substantial effect. This may explain the lack of significant differences found in the current experiment.

The results also reveal that rendering with individualised HRTFs led to significantly higher realism and externalisation compared to non-individualised renderings when there were no head movements. Conversely, with head movements both renderings induce similar ratings. This result aligns with findings by Starz et al. \cite{starz2025comparison}, who reported that non-individual HRTFs achieved plausibility ratings comparable to individual HRTFs in a scenario where visual cues were provided via a head-mounted display and head-tracking applied. Moreover, Hendrickx et al. and Brimijoin et al. \cite{hendrickx2017influence, brimijoin2013contribution} suggested that head movements might be more beneficial for externalisation when binaural synthesis is not individualised. Therefore, in this study, the non-individualised rendering might have benefited from head movements, producing similar levels of externalisation to the individualised rendering in the session with head movements. Considering the strong correlation between realism and externalisation, this may have ultimately affected the realism rating of the non-individualised rendering as well.

This study suggests that the individualisation of HRTFs does not enhance realism and externalisation when head movements are available. However, several characteristics of the protocol must be underlined as they could significantly bias the perceptual differences between individualised and non-individualised renderings. 

\subsection{Study limitations and future work}

First, the participant sample was relatively small and not naïve to spatial audio concepts. Although measures such as withholding hypotheses, randomising presentation order, anonymising conditions, and using performance-based localisation tasks were implemented to mitigate bias, subtle expectancy effects cannot be fully excluded. The limited sample size also reduces statistical power and may increase sensitivity to sampling variability, with some results therefore requiring confirmation with a larger sample size. Moreover, no pilot study was conducted, meaning that certain design aspects—such as the range of stimulus types—were not fully optimised in advance. In retrospect, the average session length (around 30 minutes) would have permitted the inclusion of additional stimuli, such as noise, without exceeding practical limits. The stimulus type was found to be more critical than HRTFs individualisation on stimulus plausibility \cite{starz2025comparison}. Future work should therefore recruit larger, more naïve listener pools, explicitly assess the role of prior experience, and refine the stimulus set to strengthen the robustness and generalisability of the findings.

Second, the localisation method required discrete responses rather than continuous estimates of source direction. This precluded the use of standard localisation metrics (polar, azimuth, elevation error) and limited comparability with studies using continuous tasks. However, discrete formats reduce response variability and noise, thus improving within-experiment robustness at the cost of generalisability.

The findings suggest that HRTF individualisation is beneficial even in dynamic conditions, specifically for localisation performance. However, these results must be interpreted in light of specific aspects of the protocol. The experiment was conducted in a semi-anechoic room. The lack of reverberation is known to potentially increase the perceptual differences between HRTFs when compared to more reverberant rooms \cite{rummukainen2021head}. Additionally, all virtual source locations were associated with a loudspeaker, providing participants with an additional visual cue to believe the stimuli were real. This type of multisensory integration is known to improve externalisation and realism \cite{werner2016summary}. The nature of participants' head movements was not analysed in this study, investigating the strategies used by participants in utilising head movements to localise sound sources could provide additional insights into why certain source positions benefited from HRTF individualisation.

Although the estimated end-to-end latency of the head-tracking and rendering pipeline (15.3 ms) is below reported detectability thresholds \cite{brungart2005detectability} and within the range generally considered acceptable for interactive binaural rendering \cite{best2020sound,brimijoin2013contribution}, it is still possible that even sub-threshold latencies contributed to small degradations in externalisation. Such effects may be particularly relevant in highly dynamic listening scenarios, where latency can interact with other factors such as spectral cues or room simulation accuracy.

A further limitation lies in the lack of a standardised baseline for evaluating the realism of virtual sound sources. The literature demonstrates substantial contradictions in reported outcomes, with results varying not only according to experimental conditions, such as stimulus type, source location, and room acoustics, but also, and perhaps more critically, according to the evaluation paradigm and reporting method. For example, studies using plausibility ratings \cite{starz2025comparison} may reach contradictory conclusions to those using authenticity-based direct comparison tests \cite{brinkmann2017authenticity}, even when investigating similar rendering conditions. Our study is no exception: the choice of plausibility-based ratings likely influenced the observed outcomes and may differ from results obtained with stricter or alternative evaluation methods. This variability highlights the need for establishing an accepted baseline framework for realism evaluation in spatial audio research in AAR.

Future research should also explore the full implications of listener movements on the type and resolution of the binaural rendering process used in AAR scenarios, incorporating six degrees of freedom tracking and other conditions related to the rendering, including HRTFs and reverberation. Additionally, the study's use of a semi-anechoic room should be extended to environments with varying reverberations, to investigate the interaction between reverberation and HRTF reproduction.

\section{Conclusion}\label{sec13}

This study investigates whether individual HRTFs are essential for the localisation and realism of virtual sound sources, considering the presence or absence of head movements. These criteria were examined using binaural reproduction of virtual sound sources with either a KEMAR HRTF or an individual HRTF, compared to reproduction through loudspeakers. The aim was to emulate a simple Audio Augmented Reality (AAR) scenario where the virtual sound sources are associated with real visual objects.

It was initially hypothesised that the beneficial effect of head movements on localisation would surpass the benefits provided by individualised HRTFs. This hypothesis was based on the work of Begault et al. \cite{begault2001direct}, and the protocol used in this study was similar to theirs. The results revealed different outcomes: the benefits of HRTF individualisation on localisation were significant only when head movements were possible. This was possibly due to differences between the two studies in terms of the direction of appearance of the sound sources. In this study, sources were present on a sphere surrounding the participants, while in Begault et al. sources were only present in the horizontal plane.

It was also hypothesised that allowing head movements might reduce the perceived realism of virtual sound sources, as it becomes more apparent that the sound is originating from headphones rather than from a loudspeaker. This reduction in realism would, in turn, diminish the benefit of HRTF individualisation. This effect was indeed observed in the results, as head movements seemed to diminish the perceptual differences between the individualised and non-individualised HRTF renderings. As a result, the benefits of HRTF individualisation for enhancing realism were significant only when head movements were absent.

\section*{Acknowledgements}

This project has received funding from the SONICOM project (www.sonicom.eu), a European Union’s Horizon 2020 research and innovation programme under grant agreement No. 101017743.





\bibliography{sn-bibliography.bib}

\end{document}